%% file: main.tex
\documentclass[conference]{IEEEtran}
\IEEEoverridecommandlockouts%
\usepackage{cite}
\usepackage[para]{footmisc}
\usepackage{amsmath,amssymb,amsfonts}
\usepackage{algorithmic}
\usepackage{enumitem}
\usepackage{graphicx}
\graphicspath{{./img/}}
\usepackage{textcomp}
\usepackage[table,xcdraw]{xcolor}
\usepackage{diagbox}
\usepackage{colortbl}
\usepackage{todonotes}
\usepackage{multirow}
\usepackage{url}
\usepackage{nohyperref} 
\usepackage[binary-units,per-mode=symbol]{siunitx}
\usepackage{subcaption}
\usepackage[nohyperlinks,nolist]{acronym} 
\usepackage[]{algorithm2e}
\usepackage{tikz}
\usepackage{booktabs}
\usepackage[absolute,showboxes]{textpos}
\usetikzlibrary{patterns}
\usetikzlibrary{spy}
\usepackage{pgfplots}
\pgfplotsset{compat=newest}
\usepgfplotslibrary{units}
\usepgfplotslibrary{groupplots}
\makeatletter
\pgfplotsset{
 unit code/.code 2 args=
   \begingroup
   \protected@edef\x{\endgroup\si{#2}}\x
}
\makeatother
\usetikzlibrary{matrix}

\definecolor{CoreGray}{HTML}{BFBFBF}
\definecolor{CoreBlack}{HTML}{333333}
\definecolor{CoreBlue}{HTML}{002E7D}
\definecolor{CoreGreen}{HTML}{6AAC8E}
\definecolor{CoreRed}{HTML}{C80000}
\definecolor{CoreYellow}{HTML}{E6AC00}
\definecolor{CoreWhite}{HTML}{FFFFFF}
\colorlet{LightCoreGray}{CoreGray!20}
\colorlet{LightCoreBlack}{CoreBlack!20}
\colorlet{LightCoreBlue}{CoreBlue!20}
\colorlet{LightCoreGreen}{CoreGreen!30}
\colorlet{LightCoreRed}{CoreRed!20}
\colorlet{LightCoreYellow}{CoreYellow!20}
\colorlet{LightCoreWhite}{CoreWhite!20}



\begin{document}

\title{
Strategies for Integrating Control Flows in Software-Defined
In-Vehicle Networks and Their Impact on Network Security
\thanks{This work is funded by the German Federal Ministry of Education and Research (BMBF) within the SecVI project.}
}
\author{\IEEEauthorblockN{Timo H\"ackel, Anja Schmidt, Philipp Meyer, Franz Korf, and Thomas C. Schmidt}
\IEEEauthorblockA{\href{http://www.haw-hamburg.de/ti-i}{\textit{Dept. Computer Science}},
\href{http://www.haw-hamburg.de/ti-i}{\textit{Hamburg University of Applied Sciences}}, Germany \\
\{\href{mailto:timo.haeckel@haw-hamburg.de}{timo.haeckel}, \href{mailto:anja.schmidt@haw-hamburg.de}{anja.schmidt}, \href{mailto:philipp.meyer@haw-hamburg.de}{philipp.meyer}, \href{mailto:franz.korf@haw-hamburg.de}{franz.korf}, \href{mailto:t.schmidt@haw-hamburg.de}{t.schmidt}\}@haw-hamburg.de}
}

\maketitle

\setlength{\TPHorizModule}{\paperwidth}
\setlength{\TPVertModule}{\paperheight}
\TPMargin{5pt}
\begin{textblock}{0.8}(0.1,0.02)
     \noindent
     \footnotesize
     If you cite this paper, please use the original reference:
     Timo H\"ackel, Anja Schmidt, Philipp Meyer, Franz Korf, and Thomas C. Schmidt. Strategies for Integrating Control Flows in Software-Defined In-Vehicle Networks and Their Impact on Network Security. In: \emph{2020 IEEE Vehicular Networking Conference (VNC) (IEEE VNC 2020)}. IEEE Press, December 2020.
\end{textblock}

\begin{abstract}
    Current \acp*{IVN} connect \acp*{ECU} via domain buses.
    A gateway forwards messages between these domains.
    Automotive Ethernet emerges as a flat, high-speed backbone technology for IVNs that carries the various control flows within Ethernet frames.
    Recently, \ac*{SDN} has been identified as a useful building block of the vehicular domain, as it allows the differentiation of packets based on all header fields and thus can isolate unrelated control flows.
    
    In this work, we systematically explore the different strategies for integrating automotive control flows in switched Ether-networks and analyze their security impact for a software-defined IVN.
    We discuss how control flow identifiers can be embedded on different layers resulting in a range of solutions from fully exposed embedding to deep encapsulation.
    We evaluate these strategies in a realistic IVN based on the communication matrix of a production grade vehicle, which we map into a modern Ethernet topology.
    We find that visibility of automotive control flows  within packet headers is essential for the network infrastructure to enable isolation and access control.
    With an exposed embedding, the SDN backbone can establish and survey trust zones within the IVN and largely reduce the attack surface of connected cars.
    An exposed embedding strategy also minimizes communication expenses.
\end{abstract}


\input{tex/0_acronyms}

\input{tex/1_intro}
\input{tex/2_background}
\input{tex/3_concept}
\input{tex/4_evaluation}
\input{tex/5_conclusion}

\bibliographystyle{IEEEtran}
\bibliography{HTML-Export/all_generated,rfcs,bib/special}

\end{document}

%% file: tex/0_acronyms.tex
\begin{acronym}
	\acro{ACC}[ACC]{Adaptive Cruise Control}
	\acro{ADS}[ADS]{Anomaly Detection System}
	\acroplural{ADS}[ADSs]{Anomaly Detection Systems}
	\acro{API}[API]{Application Programming Interface}
	\acro{AVB}[AVB]{Audio Video Bridging}
	\acro{ARP}[ARP]{Address Resolution Protocol}
	\acro{BE}[BE]{Best-Effort}
	\acro{CAN}[CAN]{Controller Area Network}
	\acro{CBM}[CBM]{Credit Based Metering}
	\acro{CBS}[CBS]{Credit Based Shaping}
	\acro{CMI}[CMI]{Class Measurement Interval}
	\acro{CoRE}[CoRE]{Communication over Realtime Ethernet}
	\acro{CT}[CT]{Cross Traffic}
	\acro{CM}[CM]{Communication Matrix}
	\acro{DoS}[DoS]{Denial of Service}
	\acro{DPI}[DPI]{Deep Packet Inspection}
	\acro{ECU}[ECU]{Electronic Control Unit}
	\acroplural{ECU}[ECUs]{Electronic Control Units}
	\acro{HTTP}[HTTP]{Hypertext Transfer Protocol}
	\acro{HMI}[HMI]{Human-Machine Interface}
	\acro{IA}[IA]{Industrial Automation}
	\acro{IDS}[IDS]{Intrusion Detection System}
	\acroplural{IDS}[IDSs]{Intrusion Detection Systems}
	\acro{IEEE}[IEEE]{Institute of Electrical and Electronics Engineers}
	\acro{IoT}[IoT]{Internet of Things}
	\acro{IP}[IP]{Internet Protocol}
	\acro{ICT}[ICT]{Information and Communication Technology}
	\acro{IVNg}[IVN]{In-Vehicle Networking}
	\acro{IVN}[IVN]{In-Vehicle Network}
	\acroplural{IVN}[IVNs]{In-Vehicle Networks}
	\acro{LIN}[LIN]{Local Interconnect Network}
	\acro{MOST}[MOST]{Media Oriented System Transport}
	\acro{NADS}[NADS]{Network Anomaly Detection System}
	\acroplural{NADS}[NADSs]{Network Anomaly Detection Systems}
	\acro{OEM}[OEM]{Original Equipment Manufacturer}
	\acro{RC}[RC]{Rate-Constrained}
	\acro{REST}[ReST]{Representational State Transfer}
	\acro{SDN}[SDN]{Software-Defined Networking}
	\acro{SOA}[SOA]{Service-Oriented Architecture}
	\acro{SOME/IP}[SOME/IP]{Scalable service-Oriented MiddlewarE over IP}
	\acro{SR}[SR]{Stream Reservation}
	\acro{SRP}[SRP]{Stream Reservation Protocol}
	\acro{SW}[SW]{Switch}
	\acroplural{SW}[SW]{Switches}
	\acro{TAS}[TAS]{Time-Aware Shaping}
	\acro{TCP}[TCP]{Transmission Control Protocol}
	\acro{TDMA}[TDMA]{Time Division Multiple Access}
	\acro{TSN}[TSN]{Time-Sensitive Networking}
	\acro{TSSDN}[TSSDN]{Time-Sensitive Software-Defined Networking}
	\acro{TT}[TT]{Time-Triggered}
	\acro{TTE}[TTE]{Time-Triggered Ethernet}
	\acro{UDP}[UDP]{User Datagram Protocol}
	\acro{QoS}[QoS]{Quality-of-Service}
	\acro{V2X}[V2X]{Vehicle-to-X}
	\acro{WS}[WS]{Web Services}
	\acro{ZC}[ZC]{Zone Controller}

\end{acronym}

%% file: tex/1_intro.tex

\section{Introduction}%
\label{sec:introduction}

\acp{IVN} connect sensors, actuators, and \acp{ECU} which are traditionally attached to bus systems such as the \ac{CAN}.
In recent years, automotive Ethernet has emerged as the high bandwidth communication technology for in-car backbones, which will soon form switched, flat topologies~\cite{brkw-aeaJR-17}.
At the network edge, local gateways will then embed control communication of buses into Ethernet frames.

Traditionally, \acp{IVN} are organized in domains such as infotainment, powertrain, and comfort and each \ac{ECU} is part of one domain.
The communication matrix of a vehicle specifies all communication relations between \acp{ECU}.
A common approach to bring those control flows onto the Ethernet backbone is by encapsulating bus messages into upper protocol layers such as SOME/IP---a tunneling approach.
Residing in an application layer protocol, the network neither can identify the control messages nor is it aware of the actual endpoints.
With increasingly interconnected \acp{ECU} this can open new attack vectors as every gateway participating in the tunneling layer is able to send and receive any message in that tunnel.
In contrast, control messages can be directly embedded in Ethernet frames using multicast destination addresses specific to message types and domain specific VLAN IDs.
This keeps communication contexts exposed to the switching infrastructure and each vehicular control flow can be precisely identified.

\acf{SDN} was first introduced in campus networks~\cite{mabpp-oeiJR-08} and promises to reduce the complexity of network control while increasing adaptability.
In \ac{SDN}, a central controller manages the control plane of the network.
Network devices decide on the forwarding of packets based on a programmable matching pipeline managed by the controller via the OpenFlow protocol.
The OpenFlow forwarding pipeline identifies network flows based on header information from layer~2 to layer~4.
In recent years, use cases for SDN extended to the vehicular domain~\cite{hmg-rsarn-18}.
The \ac{SDN} matching pipeline promises to be a powerful tool that can precisely identify and separate control flows in the network. 

In this work, we contribute strategies to integrate control flows in a software-defined in-vehicle Ethernet backbone and analyze their impact on the security and performance of the \ac{IVN}.
We compare two hidden tunneling approaches and an exposed layer 2 embedding to show how identifiers of a control flow can be embedded at different layers.
We evaluate the integration strategies in a realistic \ac{IVN} based on the \ac{CAN} network of a production car, which we transformed into a modern SDN-controlled Ethernet topology.

The remainder of this work is structured as follows.
Section~\ref{sec:background_&_related_work} reviews automotive Ethernet and related work. 
Section~\ref{sec:desing} illustrates the design space for embedding control flows in software-defined \acp{IVN}. 
Security and performance evaluations of the different integration strategies are shown in Section~\ref{sec:eval}, while Section~\ref{sec:case_study} takes a closer look on a critical control flow.
Finally, Section~\ref{sec:conclusion_&_outlook} concludes with an outlook. 

%% file: tex/2_background.tex

\section{Background and Related Work}%
\label{sec:background_&_related_work}
Modern cars implement functions using sensors, actuators, and \acp{ECU} that are linked via a combination of traditional bus systems and Ethernet technologies for newly selected links.
We use \ac{CAN} control flows as an example to discuss the integration of traditional bus communication into Ethernet.

In-vehicle control communication is precisely defined, and the content of messages is known.
All installed \acp{ECU} and their communication relations are specified in the communication matrix of a vehicle.
Each control flow in the matrix has exactly one sender, a list of receivers, and a CAN ID which is a unique vehicle wide identifier.
Each \ac{ECU} belongs to one domain (e.g., drivetrain) and the domain of a control flow is determined based on its sending \ac{ECU}.

\acp{ECU} in current vehicles are grouped in functional domains and interconnected by proprietary bus systems. 
These domains often physically extend over large regions of the car. 
Functions based on information from different domains require messages transferred from one domain to another via a central gateway.
The stepwise transition to flat Ethernet introduces processing clusters that are connected via a switched Ethernet backbone. Clusters bundle virtual functions and perform computation-intensive tasks.
In this work, we use a zone topology~\cite{brkw-aeaJR-17} which connects all legacy \acp{ECU} to a zone controller in their physical vicinity within the vehicle (e.g., front left).

In such architectures, it is essential that \ac{QoS} guarantees are provided by the network to control concurrent mixed-critical communication. 
\acf{TSN} is a set of standards (see~IEEE 802.1Q-2018) which extend Ethernet with the ability to forward real-time- and cross-traffic concurrently and is the leading candidate for deployment in the vehicular world.

Existing \ac{CAN}-Ethernet gateway solutions embed CAN traffic in application layer protocols.
Kern et al.~\cite{krst-gseJR-11} compare aggregation strategies for automotive CAN-Ethernet gateways.
For all strategies they send all messages of one CAN bus via one UDP/IP flow.
\mbox{AUTOSAR} specifies \ac{SOME/IP} to transport control information over Ethernet~\cite{a-somip-16} and recommends embedding the traditional bus communication using the full SOME/IP header.
In contrast, Herber et al.~\cite{hrwh-rtcJR-15} evaluate sending CAN frames over AVB streams, which are now integrated in \ac{TSN}.
None of these approaches exposes context information about the control flows in networking protocols that are used for the forwarding decision in the network.
Instead, Rumez et al.~\cite{rdgks-iabJR-19} integrate access control in automotive gateways that shall  filter unwanted messages based on policies and attributes. 

Wang et al.~\cite{wlk-ncaJR-19} discuss network and communication technologies for both in-vehicle and \ac{V2X} communication in the era of autonomous driving. 
They argue that more flexibility will be necessary to manage the growing dynamics of \acp{IVN}.
Connected cars require new standards for cyber-security, such as specified in SAE/ISO 21434.
Hu et al.~\cite{hl-rscJR-18} review secure communication approaches for \acp{IVN} including message authentication, encryption, and intrusion protection. The authors advise for the traditional protective functions such as intrusion detection, firewalling, and access control at the network transition between trust zones.
We will show that the network itself can help establishing those trust zones via admission control and network separation.

\acf{SDN} promises to reduce the complexity of networks and increase the adaptability~\cite{krvra-sdncs-15}.
Shin et al.~\cite{sxhg-ensJR-16} review network security enhancements by SDN, for instance using  central dynamic access control, network separation and firewalls.
Halba et al.~\cite{hmg-rsarn-18} show how \ac{SDN} can improve safety and robustness of \acp{IVN}.
In previous work, we examined the combination of \ac{SDN} and \ac{TSN} for in car networks~\cite{hmks-snsti-19}.
These investigations promised that \ac{SDN} can add value to the resilience, security, and adaptivity of the automotive environment, while its control overhead can be mitigated without a delay penalty in real-time communication.

Evaluating network security mechanisms is difficult as it is hard to predict the risk of unknown vulnerabilities~\cite{wjscn-kzdJR-14}.
Ruddle et al.~\cite{rwwir-sraJR-09} present a concept for security analysis of automotive on-board networks based on dark side attack scenarios.
Fundamental work on the automotive attack surface has been done by Checkoway et al.~\cite{cmkas-ceaas-11}.
Miller and Valasek~\cite{mv-sraJR-14} analyzed the remote attack surfaces of many passenger vehicles.
In later work~\cite{mv-reuJR-15}, they outlined a remote attack against an unaltered 2014 Jeep Cherokee.
Longari et al.~\cite{lccz-sdfJR-19} present a framework for automotive on-board network risk analysis.
Based on an attack tree built from real-world incidents they evaluate how network topologies can be hardened by introducing additional gateways. 
Our work concentrates on reducing the attack surface by monitoring and separating network communication flows already on the MAC layer.

%% file: tex/3_concept.tex

\section{Design Space for Embedding Control Flows}
\label{sec:desing}
A \textbf{Control Flow (CF)} is a sequence of messages with the same identifier in the vehicle communication matrix.
Sent from a single origin, it follows a point to multipoint relation to reach one or multiple receivers. 
The source embeds the CF identifier into packet headers, which allows receivers to identify the message context.

A \textbf{Network Flow (NF)} is a sequence of related packets transported from a particular source to a particular destination in the network.
Using multicast, a NF can reach multiple sinks simultaneously. 
Multicast is a common use case in \acp{IVN} since the same information is often required at different \acp{ECU}.

Network devices identify a NF by matching packet header fields. 
In addition of layer~2 addressing, Ethernet switches may consider VLAN tags.
Routers forward according to layer~3 addresses.
The OpenFlow pipeline in SDN switches allows to match all header fields from layer~2 to layer~4. 
Some SDN vendors include extensions to the OpenFlow protocol which allow matching specific bytes within the transport payload.
In this work, we assume that the network consists of \ac{SDN} forwarding devices able to precisely match all header fields from 
layer~2 to layer~4.

\subsection{Embedding Strategies for Control Flow Information} 

\label{sec:layer_embeddings}
\begin{figure}
    \centering
    \begin{subfigure}[c]{\linewidth}
        \centering
        \includegraphics[width=\linewidth, 
        trim=0.62cm 0.62cm 0.62cm 0.62cm, clip=true
        ] {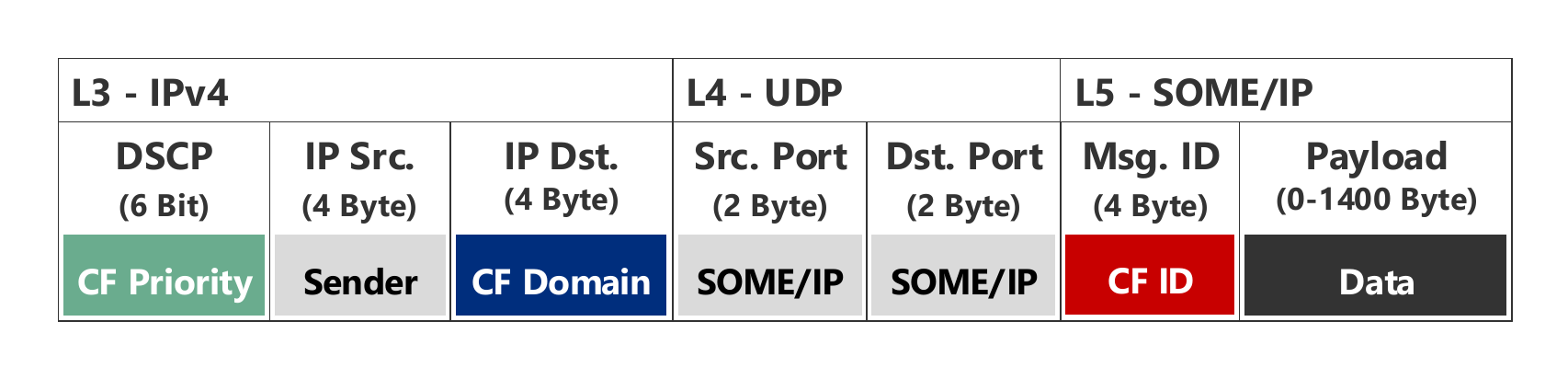}
        \caption{Hidden SOME/IP tunnel embedding}
        \label{fig:embeddingsomeip}
    \end{subfigure}    
    \begin{subfigure}[c]{\linewidth}
        \centering
        \includegraphics[width=\linewidth, 
        trim=0.62cm 0.62cm 0.62cm 0.62cm, clip=true
        ] {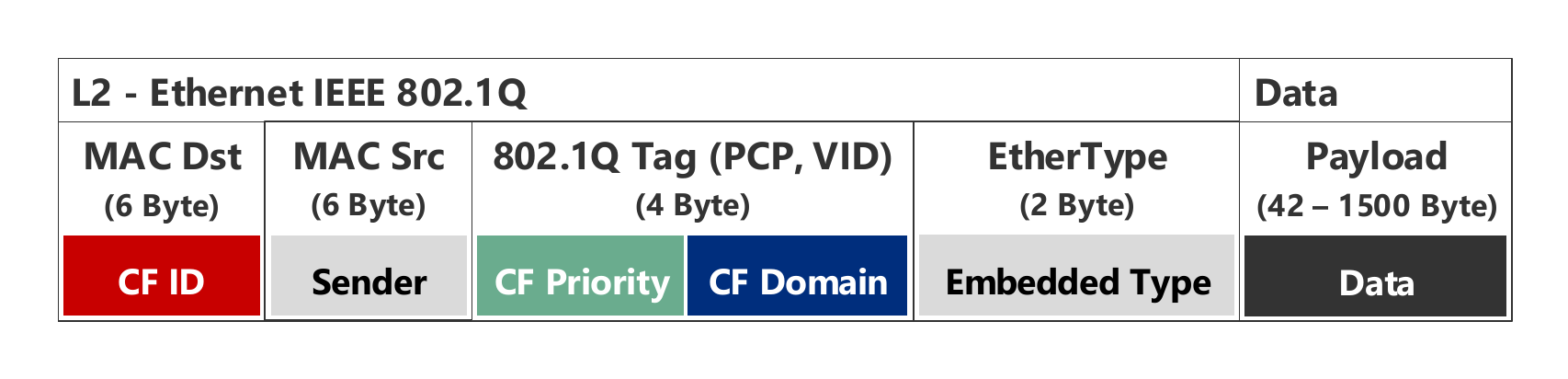}
        \caption{Exposed Ethernet embedding}
        \label{fig:embeddingl2}
    \end{subfigure}    
    \caption{
        Strategies for embedding Control Flow (CF) context  
        }
    \label{fig:embeddings}
    \vspace{-10pt}
\end{figure}

CAN-Ethernet \textbf{gateways} exchange data between a CAN Bus and an Ethernet network.
For this purpose, the CF is repackaged between CAN and Ethernet frames.
Figure~\ref{fig:embeddings} presents two examples for CF embedding on different layers. 
Each CF has an identifier, a sender, a priority, and a domain. 
Depending on the repackaging, the context of the traffic can be \textbf{hidden} or \textbf{exposed} for the network.

Hidden embeddings are a common approach for repackaging in \acp{IVN} which \textbf{tunnel} the data using an application layer protocol.
Figure~\ref{fig:embeddingsomeip} shows a hidden embedding that uses SOME/IP as an application layer tunneling protocol.
The CF id is embedded in the message ID field of the SOME/IP header and the data is embedded as a payload. 
The example uses one UDP socket per domain with reserved UDP ports for SOME/IP traffic.
The IPv4 destination multicast address contains the domain identifier. 
The Differentiated Services Code Point (DSCP) is filled with the CF priority.
The Ethernet header is automatically generated by the IP stack.
The stack can be configured to map a certain DSCP to a PCP for \ac{QoS} on layer~2. 
All CFs of one domain are sent to the same destination address, which creates a domain tunnel.
The resulting NF is identified based on the sending node and the multicast destination.
The destinations of the NF are the joint group of the receivers of the CFs in the tunnel. 

In contrast, the repackaging can be done in a fully exposed manner. 
The gateway embeds the context of the CF in the packet headers, which are used for the forwarding decision in the network.
Figure~\ref{fig:embeddingl2} shows an exposed embedding in the Ethernet header.
It uses a raw Ethernet socket to have full control on what context information about the message is encoded on layer~2.
The CF ID is encoded in the destination MAC as a multicast address.
The PCP in the 802.1Q tag is used to embed the CF priority which enables \ac{QoS}.
A different virtual lan is created for each domain to strictly separate bus domains while maintaining the control over each message.
We embed the data of the message inside the layer~2 payload.
The EtherType field contains information about the embedded data type, such as a custom EtherType for CAN data frame embedding.
The resulting NF identifies exactly one vehicle CF based on the multicast destination MAC address.
The source and destinations of the NF are equal to sender and receivers of the transported CF.

\subsection{Separating In-Vehicle Control Flows}
\label{sec:concept}

The embedded context information inside the packet header fields that are used for the forwarding decision enables the network to identify which CF is transported via which NF.
In this work, we compare three integration approaches for control traffic in Ethernet networks which either use a fully exposed layer~2 embedding or a hidden SOME/IP tunnel embedding. 
The integration approach is the key factor in how precisely CFs can be separated.

\textbf{Separation by message} performs a fully exposed layer~2 embedding that is used for the forwarding decision in the network.
This way, each CF has its own point to multipoint NF that can be identified and separated in the network by matching the multicast destination MAC address.

\textbf{Separation by domain} uses the hidden SOME/IP tunneling approach and encodes a domain identifier in the destination IP address. 
Those domains are already in use in current vehicles for creating domain buses.
The domain of a CF is selected based on the domain of its sender \ac{ECU}.
For each sender there is one point to multipoint NF, tunneling all CFs of one domain.
The NFs can be distinguished from others by matching the source IP and multicast destination IP.

\textbf{Separation by topic} uses the hidden SOME/IP tunneling approach as well.
Topics are groups of similar CFs. 
In full CAN bus architectures, each \ac{ECU} belongs exactly to one domain bus.
Introducing a new domain for CFs also means creating a new physical domain bus.
On an Ethernet backbone, a new topic only requires a new tunnel on an existing physical link.
With increasing amounts of cross-domain communication, it seems to be a good idea to introduce new fine-grained topics for CFs.
Smaller groups of CFs than the original vehicle domains can thus be formed on an Ethernet backbone without the need for additional wiring.
For this work, we assign custom topic identifiers to each CF in the communication matrix.
We use the same topic identifier for CFs with a common context.
For example, all CFs carrying motor control information are assigned to the motor control topic and all CFs regarding light control are assigned the light control topic. 
There are some CFs we could not group into topics.
For those we create a separate topic, which only contains the one CF.
The topic identifier is encoded in the multicast IP address.
Each topic has one point to multipoint NF for each sender of the topic.

\subsection{Impact on In-Vehicle Network Design}
\label{sec:network_design}
Higher layer embeddings allow the comfort of using the full IP stack.
If the CF information is embedded on the application layer it cannot be used for forwarding decisions without violation of the OSI model.
Layer~2 information is only valid in the local \ac{IVN}, so Ethernet embeddings that do not use the internet protocol are not routable. 
This reduces the attack surface of in-vehicle CFs. 

\begin{figure*}
    \centering    
    \includegraphics[width=0.85\linewidth, trim=0.62cm 0.62cm 0.62cm 0.62cm, clip=true] {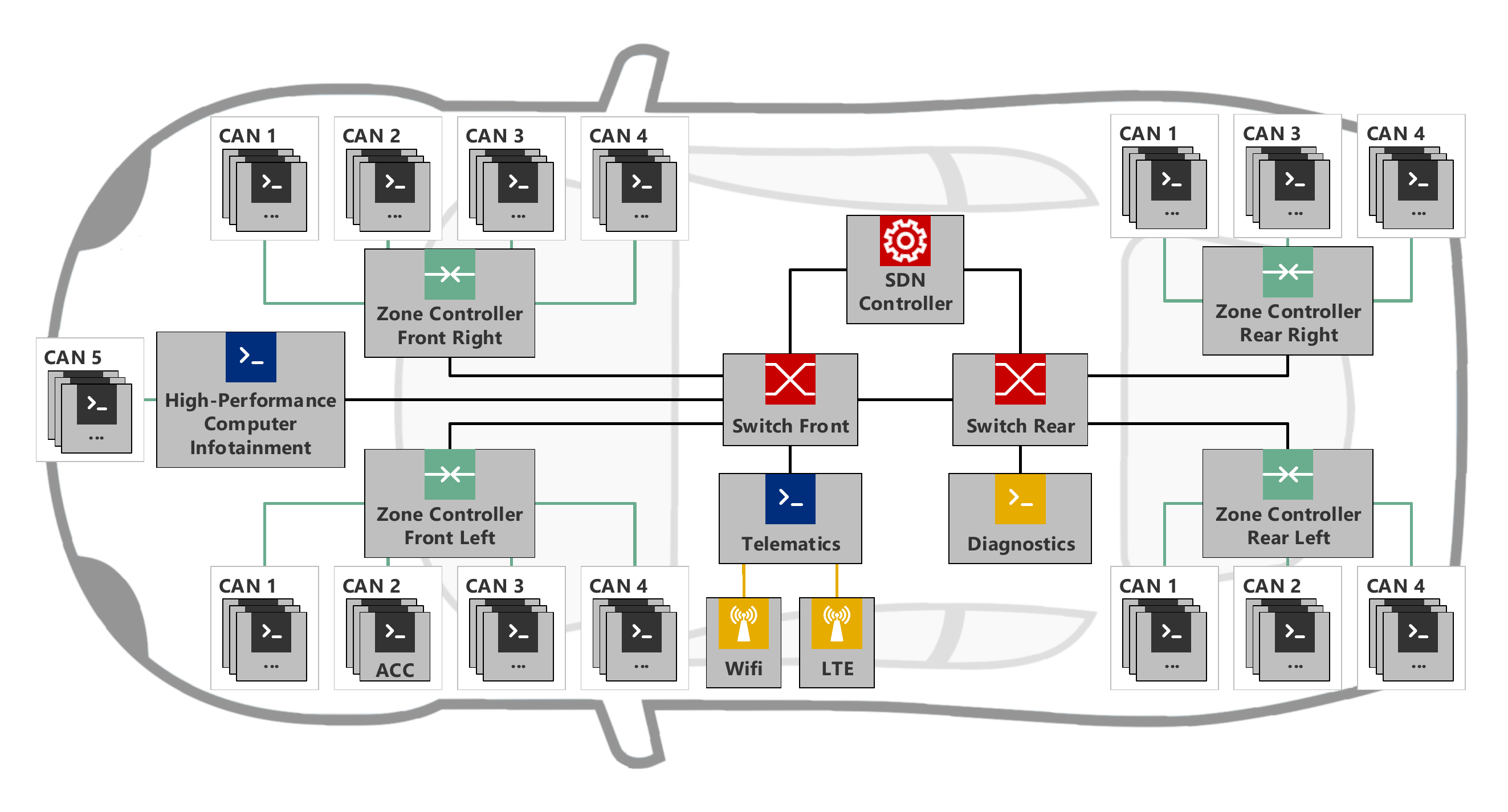}
    \includegraphics[width=0.7\linewidth, trim=0.62cm 0.62cm 0.62cm 0.62cm, clip=true] {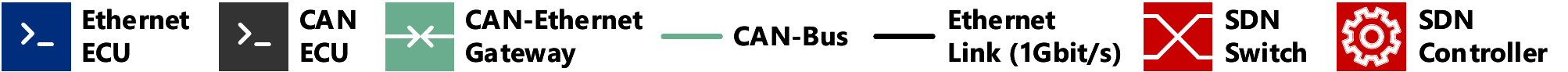}
    \caption{\ac{IVN} of a production vehicle transformed to a software-defined Ethernet network in a zone topology}
    \label{fig:topo}
    \vspace{-7pt}
\end{figure*}

\subsubsection{Message aggregation}
Aggregation is a concept to reduce overhead when relatively small control information is embedded into Ethernet. 
For example, a CAN message has 8 Byte payload while an Ethernet frame has a minimal size of 64 Bytes which results in 42 byte minimum payload. 
Kern et al. compare aggregation strategies including one-to-one mapping and buffering~\cite{krst-gseJR-11}.
Buffered approaches delay messages for a certain period before they are sent alone or together with other messages.
The one-to-one strategy sends each message in its own Ethernet packet.
The results of Kern et al. show that this leads to around 4 times higher bandwidth usage compared to buffered approaches. 
On the other hand, the latency and jitter are increased by buffering.
With different embedding concepts aggregation of control information might be hindered.
Domain tunneling allows aggregation of all messages of one domain which is the most common approach to save bandwidth. 
The finer the separation of CFs, the less useful aggregation of frames becomes as messages are delayed and only a small amount of bandwidth is saved.

\subsubsection{IP fragmentation}
Packets that are larger than the maximum transfer unit of an Ethernet frame (1500 Byte) are split into multiple fragments.
Only one fragment contains the transport header.
Therefore, transport header fields can only be used for forwarding decisions in SDN when there is no fragmentation.
Fragment reassembly in forwarding device is not appropriate in a real-time system.

\subsubsection{TSN configuration}
Control information is often transmitted in cycles and with similar data size.
\ac{TSN} allows to control the timing of NFs by reserving bandwidth or timeslots.
The reservation is more precise the better CFs can be differentiated in the network.
For exposed embeddings, the timing behavior can be determined much more precisely than for hidden embeddings, which results in less reservation overhead.
On the other hand, the planning effort for time-slot calculation increases exponentially with an increasing number of small timeslots.
Aggregation of messages makes it more difficult to calculate the timings in the network because of bursts, and varying size and period of packets.

\subsubsection{Security mechanisms}
An exposed embedding can help other security concepts such as network anomaly detection systems or deep packet inspection as they can monitor all messages of specific CFs.
Encryption might influence embedding options.
The layers used for the forwarding decision should not be encrypted, e.g. the IP address cannot be used when the frame is encrypted with MACsec.
On the other hand, encryption might be weakened as the context of the message is visible for every receiver of the NF.

\subsubsection{Device integration}
If vehicles are upgraded, new functions might require the installation of new \acp{ECU}.
For separation by domain or topic, new devices can join the existing tunnels.
This will enable them to send and receive all CFs of this tunnel.
With separation by message, new devices need to join each CF to receive them.
Also, new CFs might be introduced to the network.
For tunneling approaches, new CFs only require a new NF if they have a new sender or belong to a new domain/topic.
With separation by message, for each new CF a new NF must be created.

%% file: tex/4_evaluation.tex

\section{Evaluation}
\label{sec:eval}
We now compare the different separation concepts at hand using a realistic \ac{IVN} with real-world network traces.
First, we evaluate the network configuration generated for our topology and each separation concept.
Second, we inspect the impact of the separation on a specific vehicle control flow to show the effect in the case of an attack.

\subsection{Realistic In-Vehicle Network}
Figure~\ref{fig:topo} displays the \ac{IVN} used in our evaluation, which 
is derived from the \ac{CAN} network of a production car originally implemented with a central gateway architecture.
Our network is transformed into a zone topology~\cite{brkw-aeaJR-17}.
All \ac{CAN} \acp{ECU} are grouped into four zones based on their physical location in the vehicle (Front-Left (FL), Front-Right (FR), Rear-Left (RL), Rear-Right (RR)).
Each zone has one \acf{ZC} acting as a \ac{CAN}-Ethernet gateway.

The CAN bus domains are numbered from 1 to 5.
Despite the placement of the \ac{CAN}-\acp{ECU}, the original domain buses are retained, e.g., CAN 1 on ZC FL and CAN 1 on ZC FR were originally one domain bus.
A High-Performance Computer (HPC) handles the Infotainment~(I) system and acts as a gateway between all CAN \acp{ECU} and the infotainment domain.
All \acp{ZC} are connected via a software-defined Ethernet backbone consisting of two switches (Front (F), Rear (R)).
An \ac{SDN} controller manages the forwarding.
Besides the original \ac{CAN} network of the production car, we introduce  \acp{ECU} on Ethernet for diagnostics and external connectivity.

We generate three different network configurations matching our separation concepts and follow the communication matrix of our original production vehicle within the network topology.
The messages are played back from recorded \ac{CAN} traces of the car.
In one configuration, \acp{ZC} embed \ac{CAN} messages fully exposed on layer~2 to enable separation by messages. 
The other two configurations embed the \ac{CAN} message in a SOME/IP tunnel and encode the topic or domain identifier in the multicast IP, enabling separation by topic or domain.
The receiving \ac{ZC} transforms the packets back to \ac{CAN} frames and forwards them to the \ac{CAN} buses where the destinations are located.

\subsection{Definition of Network Flows from Control Flows} 
The focus of our analysis is the \ac{CAN} Control Flows (CFs) that traverse the Ethernet backbone.
Therefore, we do not consider local CFs, for which the sender and all receivers are located within the same zone.
In total there are 242 different CFs forwarded via the software-defined Ethernet backbone.

\begin{table}
  \centering
  \setlength{\tabcolsep}{5pt}
  \caption{Number of Network Flows (NFs) created from integrating (the 242) Control Flows (CFs) in the backbone
  }
  \label{tab:controltonetworkflows}
  \begin{tabular}{l c c c c}
      \toprule
      \textbf{Separation} & \multirow{2}{2cm}{\textbf{\# NFs (with multiple CFs)}} 
      & \multicolumn{3}{c}{\textbf{\# CFs per NF}}
      \\ 
      & & Minimum & Average & Maximum       
      \\ \midrule
      \textbf{By Message} & 242 (0) & 1 & 1 & 1
      \\
      \textbf{By Domain} & 19 (19) & 5 & 13 & 37 
      \\
      \textbf{By Topic} & 102 (38) & 1 & 3 & 17
      \\ 
      \bottomrule
  \end{tabular}
\end{table}
\begin{figure}
	\centering
    \begin{tikzpicture}
        \begin{axis}
          [width=\linewidth,height=.6\linewidth,
          ybar,
          bar width=9pt,
          legend cell align={left},
          legend style={
           anchor=north east, font=\small},
          legend image code/.code={\draw [#1] (0cm,-0.1cm) rectangle (0.15cm,0.2cm); },
          legend columns=-1,
          xtick={1,...,4},
          xtick pos=bottom,
          xticklabels={1, 2, 3, 4},
          x tick label style={align=center},
          xmin=1,
          xmax=4,
          enlarge x limits=0.2,
          xlabel={Destinations reached [\#]},
          ylabel={Network flows [\%]}, 
          ytick = {0,20,40,60,80,100}, 
          ymin = 0,
          ymax = 105,
        ]
            \addplot[fill=CoreRed] 
            coordinates { (1,27.27) (2,16.12) (3,45.45) (4,11.16)};
            \addplot[fill=CoreYellow] 
            coordinates { (1,25.49) (2,15.69) (3,41.18) (4,17.65)};
            \addplot[fill=CoreGreen] 
            coordinates { (1,0) (2,0) (3,31.58) (4,68.42)};
            \legend{Separation by Message, by Topic, by Domain}
        \end{axis}
    \end{tikzpicture}
    \vspace{-4pt}
    \caption{
      Number of destinations reached by a share of NFs 
      }
    \label{fig:cfreachinggateways}
    \vspace{-10pt}
\end{figure}
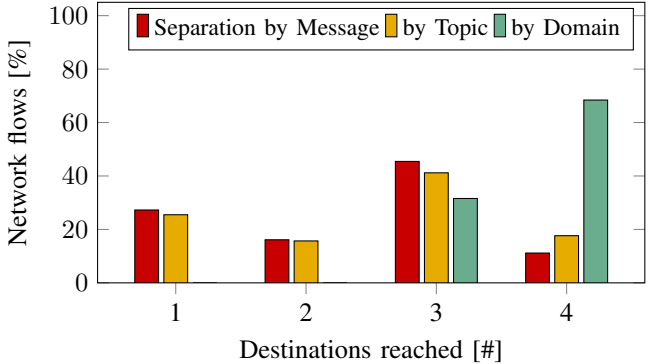 

Table~\ref{tab:controltonetworkflows} shows the generated Network Flows (NFs) for each separation concept in relation to the transported CFs.
As expected, the message separation generates 242 NFs, one for each CF.
Separation by domain generates 19 NFs, one tunnel for each sender and domain. 
All these NFs carry multiple CFs.
The topic separation generates 102 NFs, one tunnel for each sender and topic.
38 of these NFs carry multiple CFs.
This means that 74 of the topics contain only one CF and therefore behave exactly like separation by message. 
The table also shows the minimum, maximum and average number of CFs that are sent via one NF.
There is always one CF per NF using message separation. 
For domain separation, there is a minimum of 5, average of 13 and maximum of 37 CFs per NF.
So, there is at least one NF that carries 37 CFs which cannot be differentiated by the network.
As expected, separation by topic performs somewhere in between domain and message separation.
The minimum is 1 CF per NF, the average is 3 and the maximum 17.
In the following analysis, separation by message can be used as a benchmark as it implements exactly the relations of the communication matrix.

Figure~\ref{fig:cfreachinggateways} shows the share of NFs that are visible at a number of destinations not including the sender. 
The results for separation by topic and message look very similar.
For both, a quarter of all NFs reaches one gateway, and around 40\% reach fewer than three gateways.
With domain separation zero NFs reach fewer than three gateways, around a third reach exactly three and nearly 70\% reach all gateways in the network.
This is expected because the \acp{ECU} of a domain are usually distributed throughout the vehicle and therefore most domains are present in every zone.
Especially the difference in percentage of flows reaching all destinations is notable.
With topic separation 18\% of NFs reach all destinations, with separation by message only 11\%.
This indicates that the separation concept has a big influence on network overhead as CFs may be forwarded to destinations that do not need to receive them.
It is also a security issue as those receivers should not be able to receive the content.
Furthermore, multicast seems to be the best option to transport CFs in a zone topology as less than 30\% of all CFs need to be received by only one gateway.

\subsection{Impact of Control Flow Separation} 
We evaluate the separation concepts considering the following properties of CFs:
\begin{itemize}
    \item A CF is \textbf{legitimate} if the CAN source \ac{ECU} of the CF is connected to the source gateway and at least one CAN \ac{ECU} at each destination gateway is a receiver specified in the communication matrix.
    \item A CF is \textbf{received} at a destination after it originated from an \ac{ECU} connected to the source.
    \item An \textbf{oversupplied} CF is surplus and reaches the destination although it is not needed ($received \land \neg legitimate$).
    \item A CF is \textbf{permitted} if it could be sent by the source and would be forwarded to the destination. Even though not present in our communication matrix, these flows could be activated by malicious components.
\end{itemize} 

\begin{figure}
  \centering
  \begin{tikzpicture}
    \begin{groupplot}[
      group style={
        group size=3 by 1, 
        horizontal sep=0pt, 
        vertical sep=0pt,  
        yticklabels at=edge left, 
      },
      width=.45\linewidth,
      height=.75\linewidth,
      legend style={at={(-0.5,-0.29)}, anchor=north, legend columns=-1,},
      ymin=0, 
      ymax=100,
      xticklabel style={align=center,rotate=90,anchor=east},
      xtick pos=bottom, 
      symbolic x coords ={Message, Topic, Domain},
      xtick=data, 
      enlarge x limits=0.25,
    ]

      \nextgroupplot[ybar stacked, bar width=9pt, ylabel={Control flows [\%]},]
        \addplot[fill=CoreGray] coordinates {(Message,18) (Topic,18) (Domain,18)};
        \addplot[fill=CoreGreen] coordinates {(Message,0) (Topic,1) (Domain,4)};
        \addplot[fill=CoreYellow] coordinates {(Message,0) (Topic,29) (Domain,74)};
        \addplot[fill=CoreRed] coordinates {(Message,82) (Topic,52) (Domain,4)};
      
      \nextgroupplot[ybar stacked, bar width=9pt,]
        \addplot[fill=CoreGray] coordinates {(Message,14) (Topic,14) (Domain,14)};
        \addplot[fill=CoreGreen] coordinates {(Message,0) (Topic,2) (Domain,4)};
        \addplot[fill=CoreYellow] coordinates {(Message,0) (Topic,19) (Domain,49)};
        \addplot[fill=CoreRed] coordinates {(Message,86) (Topic,65) (Domain,33)};
      
        \nextgroupplot[ybar stacked, bar width=9pt,]
          \addplot[fill=CoreGray] coordinates {(Message,12) (Topic,12) (Domain,12)};
          \addplot[fill=CoreGreen] coordinates {(Message,0) (Topic,3) (Domain,7)};
          \addplot[fill=CoreYellow] coordinates {(Message,0) (Topic,16) (Domain,48)};
          \addplot[fill=CoreRed] coordinates {(Message,88) (Topic,69) (Domain,34)};
          
      \legend{
        Legitimate, 
        Oversupplied, 
        Permitted, 
        Forbidden,
      }
    \end{groupplot}
    \tikzstyle{every node}=[font=\footnotesize]
    \node[draw,shape=rectangle,above, black, fill=LightCoreYellow, align=center, ] at (1.2,5.15) {Case 1: \\ZC FL $\rightarrow$ ZC FR};

    \node[draw,shape=rectangle,above, black, fill=LightCoreGreen, align=center] at (3.6,5.15) {Case 2: \\Total $\rightarrow$ ZC FR};

    \node[draw,shape=rectangle,above, black, fill=LightCoreRed, align=center] at (6.0,5.15) {Case 3: \\Total $\rightarrow$ All dest.};
  \end{tikzpicture}
  \caption{Share of legitimate, oversupplied, permitted, and forbidden CFs in the three cases marked in Table~\ref{tab:comrelations}
  }
  \label{fig:oversupply}
  \vspace{-10pt}
\end{figure}
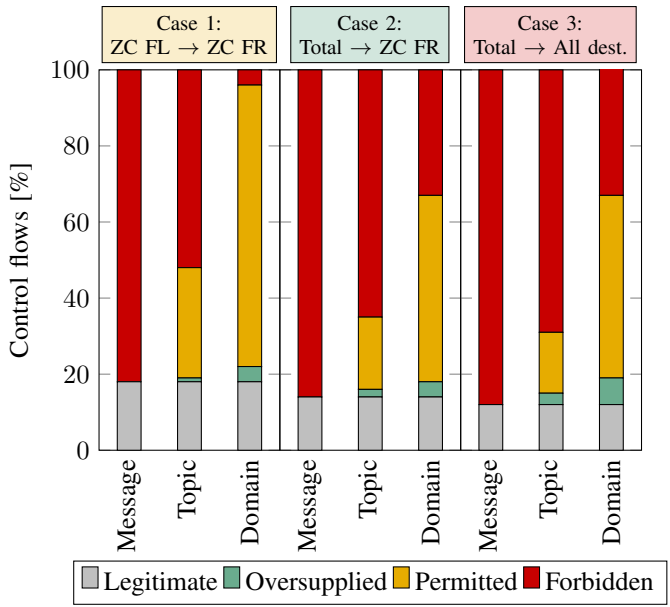

Figure~\ref{fig:oversupply} visualizes the shares of legitimate, oversupplied, permitted and forbidden CFs in relation to the maximum possible CFs in three cases. 
The first case concentrates on the CFs originating at ZC FL with the destination ZC FR, the second includes all CFs with the destination ZC FR, and the third case looks at the total CFs to all destinations.
A \textbf{forbidden} CF is filtered by the backbone and is not forwarded from the source to the destination.
All three cases clearly show the effectiveness of the strict message separation via exposed embedding.
When separated by message, the software-defined backbone only forwards the legitimate CFs and forbids any other CF.
The hidden application layer embeddings oversupply and permit illegitimate CFs.
This leads to fewer CFs that are forbidden, while still only CFs of the tunnels that are already in use by the sending nodes are allowed by the network.
More precisely, topics can improve the separation of flows in the network.
As stated earlier, we chose our topics by grouping similar messages in the communication matrix. 
There are some CFs we could not group into topics. 
This impacts the results for the separation as those CFs are perfectly separated by their own topic with only one CF.

\input{tex/4a_table.tex}

The detailed vehicle-wide analysis is exhibited in Table~\ref{tab:comrelations} showing all communication relations between the gateways.
For each destination, the CFs from all other gateways and the total of all CFs to the destination are depicted.
The percentages relate to the \textbf{maximum} CFs that are sent via the Ethernet backbone.
The three cases previously discussed and depicted in Figure~\ref{fig:oversupply} are marked accordingly.

We take a closer look at the CFs with the destination ZC FR and the source ZC FL marked as yellow.
Each gateway forwards a maximum of 242 CFs from CAN buses to other gateways via the Ethernet backbone.
There are 43 CFs from ZC FR to ZC FL specified in the communication matrix according to the placement of the \ac{CAN} \acp{ECU} in our topology.
With separation by message, the same 43 CFs (18\% of 242) are received by ZC FL from ZC FR, so the oversupply is zero.

For tunneling approaches, CFs are  forwarded to gateways that are subscribed the multicast group, even though for other CFs.
Hence with separation by domain, ZC FL receives 53 CFs from ZC FR, 10 of which are oversupplied CFs (4\% of the maximum 242).
With topic separation the oversupply is only 3 CFs. 
If a source sends one CF of a tunnel, the network also permits the sending of all other CFs of the NF.
Separation by domain permits 232 CFs from ZC FR to ZC FL, which is 96\% of all CFs.
This adds 189  CFs to the 43 legitimate.
Topic separation permits 115 CFs, which are 72 too many.
Separation by message only permits the sending of the legitimately specified 43 CFs.

We also calculate the total amount of CFs received, oversupplied, and permitted for each destination, for example ZC FR marked in green.
Since 4 sources are always considered for each destination, the maximum possible CFs to the destinations are 4 times 242, which is 968. 
The zone FR requires a total of 137 CFs.
When separated by message it receives the same 137 CFS, 158 by topic, and 177 by domain.
This results in a total of 21 CFs oversupplied to ZC FR with topic separation, and 40 with separation by domain.
For permitted flows the difference is more extreme as the hidden embeddings allow a lot more messages compared to the exposed embedding.
For topics there are 342 permitted CFs which is 205 more than legitimate.
For domain tunneling 647 CFs are permitted to the destination ZC FR which is 510 surplus CFs that can reach the destination ZC FR.

The totals for the whole network sum up all CFs between all nodes and are marked in red.
As explained earlier, NFs are point to multipoint relations.
Therefore, for 242 maximum CFs there are 582 legitimate CFs from all sources to all destinations.
For the whole network 4840 CFs would be possible as each source could send a maximum of 242 CFs to each destination. 
This is also reflected by the message separation while for topic separation 708 and for domain separation 915 CFs reach the destinations.

\subsection{Discussion} 

Gateways can filter the unneeded CFs at the cost of processing power so that no unwanted message reach a \ac{CAN} bus.
Still, the network load on the backbone increases due to the oversupplied control traffic.
In the third case, 12\% of the maximum CFs from all sources to all destinations are legitimate.
With separation by topic additional 3\% are received at the destinations, with domain separation around 7\%.
Network load on the backbone increases by 57\% with domain separation, and by 27\% when separated by topic as compared to the legitimate CFs.
These results  clearly conflict with the benefits of aggregation, which is one of the main arguments for using a hidden embedding.
Above all, it is a security issue that network nodes get exposed to messages they are not prepared for. 
This is especially the case for HPC I which receives an oversupply of 37 CFs when separated by topic and 146 CFs when separated by domain.
This is twice as many received CFs with topic separation compared to the legitimate CFs and over four times as many with domain separation. 
The separation by message is the only concept able to solve this problem with zero oversupplies.
More fine-grained tunneling could lead to better separation.
However, this approach also reduces the advantages of application layer tunneling such as easier setup with higher layer sockets, the possibility to aggregate, and fewer NFs.

From a security perspective, the CFs that a (malicious) \ac{ECU} can send are even more important than the CFs it can receive.
From all sources to all destinations, 31\% of all CFs are allowed with topic separation and even 67\% with separation by domain, while only 12\% are needed.
The domain separation with a total of 3232 allowed CFs on the network results in a 455\% increase in permitted CFs compared to the 582 legitimate CFs on the backbone.
For topic tunnels there are 1482 permitted CFs which is an increase of 155\%.
However, the difference is not as drastic for CFs permitted from HPC I, which in every separation concept can only send CFs of the infotainment domain to the other gateways.
Still, this shows the security weakness of the tunneling approach.
When a device participates in a tunnel, it can listen to all CFs of this tunnel and is permitted to send all CFs of the tunnel.

Fewer filtering of CFs eases attacks as fewer  \acp{ECU} need to be compromised to gain control over the car. 
Even if gateways filter all unwanted and illegally sent messages it would still be possible to attack the NFs, e.g., by flooding the tunnels to delay time-critical flows.
In the case of exposed embeddings, which fully isolates messages by SDN, attackers must compromise the exact sender of each CF in the Ethernet backbone to issue messages in this channel.

The presented zone topology with an enforced separation of messages by SDN is also more secure than the original network architecture of our production vehicle.
In a pure CAN bus architecture, the traffic on buses cannot be controlled.
All \acp{ECU} connected to a domain bus can send any CF and receive all messages on the bus.
A compromised \ac{ECU} can attack all other \acp{ECU} in its domain.
Gateways can use the CAN IDs to filter which messages are to be forwarded to another bus, but cannot verify the correct sender of a CF.
The zone topology splits the CAN domain buses so that less \acp{ECU} are connected to a physical bus, which allows the ZCs to filter messages and less \acp{ECU} are completely unprotected.
If each \ac{ECU} were  directly connected to the zone controller, the correct sender could be verified.
Combined with separation of messages by SDN, this could offer a perfect separation of CFs and thus a trustworthy communication backbone. 
 
\section{Case Study: Impact on a Critical Control Flow}
\label{sec:case_study}
To illustrate how CF separation can reduce the attack surface of the presented \ac{IVN}, we now pick critical CFs often used in attack scenarios from known exploitations against passenger vehicles such as~\cite{mv-reuJR-15}.
We select the acceleration request 
as a sample CF, which we examine in two different cases.
The first case assumes a compromised CAN \ac{ECU}. 
We discover all CAN \acp{ECU} that can send the acceleration request to the original recipients.
Second, an Ethernet node could be compromised. 
We discover which gateways can send the acceleration request to the original receivers via Ethernet.

The acceleration request is sent by the Adaptive Cruise Control (ACC) connected to ZC FL. 
\acp{ECU} connected to the gateways ZC FR and ZC RR require the CF.
If a compromised CAN \ac{ECU} is connected to the same physical bus as the original source or the destination of the CF, the attack cannot be prevented.
This is the case for 7 CAN \acp{ECU}. 
The gateway connected to the compromised \ac{ECU} can filter illegitimate CFs if the destinations are not on the same physical bus.
If the compromised CAN \ac{ECU} is in a different zone the CF is forwarded over the Ethernet backbone.
There is only one additional \ac{ECU} that can send the acceleration request with separation by message on the backbone.
With separation by domain, the backbone forwards the CF from 29 additional \acp{ECU} with separation by topic from 11.
Still, those would be filtered in the source and destination gateways, except the one also permitted with separation by message.
This shows the impact of the zone topology. 
Dividing the domain buses into separate zones allows the gateways and the network to filter messages.
If all \acp{ECU} were directly connected to a ZC, it could precisely identify each sender and their permitted CFs.

In case of a compromised gateway, the CFs can only be protected by the Ethernet backbone.
With separation by message the acceleration request is only received at ZC FR and ZC RR, which require the information.
Both hidden embeddings lead to oversupply of this critical CF reaching HPC~I with separation by topic, and ZC RL and HPC~I with separation by domain.
Only the original sender (ZC FL) can send the acceleration request with message separation. 
With domain and topic separation, ZC RL and ZC FR are also allowed to send the command.

The communication relations for this example show the benefits of an exposed embedding over hidden embeddings in tunneling protocols.
Hidden embeddings lead to unwanted receivers of critical CFs and permit their transmission from other network participants which is a risk for safety and security.
An exposed embedding enables the separation of all CFs in the network and can establish a trust zone between the gateways, so that only the original sending gateway is permitted to send the CF and it reaches only the necessary destinations.
This improves network security even in the case of a compromised gateway.

%% file: tex/4a_table.tex
\begin{table*}
    \centering
    \setlength{\tabcolsep}{1.5pt}
    \renewcommand{\arraystretch}{0.3}
    \caption{Communication relations between gateways in terms of the received, oversupplied, and permitted CFs for each separation concept compared to the maximum available, and legitimate CFs}
    \label{tab:comrelations}
    \begin{tabular}
      {l l c c l
        c c c l
        c c c l
        c c c }
        \toprule
        \multicolumn{4}{c}{\textbf{Networking Properties}} & &
        \multicolumn{3}{c}{\textbf{Separation by Message}} & &
        \multicolumn{3}{c}{\textbf{Separation by Topic}} & &
        \multicolumn{3}{c}{\textbf{Separation by Domain}} 
        \\\cmidrule{1-4} \cmidrule{6-8}\cmidrule{10-12}\cmidrule{14-16}
        Dest. & Src. & Maximum & Legitimate &
        & Received & Oversupplied & Permitted &
        & Received & Oversupplied & Permitted &
        & Received & Oversupplied & Permitted  
        \\ \midrule
        \rowcolor{LightCoreYellow}
        \textbf{ZC FR} & ZC FL & 242 & 43 (18\%) & &
        43 (18\%) & 0 (0\%) & 43 (18\%) & &
        46 (19\%) & 3 (1\%) & 115 (48\%) & &
        53 (22\%) & 10 (4\%) & 232 (96\%) 
        \\
        & ZC RL & 242 & 47 (19\%) &  &
        47 (19\%) & 0 (0\%) & 47 (19\%) & &
        60 (25\%) & 13 (5\%) & 115 (48\%) & &
        68 (28\%) & 21 (9\%) & 192 (79\%) 
        \\
        & ZC RR & 242 & 37 (15\%) &  &
        37 (15\%) & 0 (0\%) & 37 (15\%) & &
        42 (17\%) & 5 (2\%) & 102 (42\%) & &
        46 (19\%) & 9 (4\%) & 213 (88\%) 
        \\
        & HPC I & 242 & 10 (4\%) &  &
        10 (4\%) & 0 (0\%) & 10 (4\%) & &
        10 (4\%) & 0 (0\%) & 10 (4\%) & &
        10 (4\%) & 0 (0\%) & 10 (4\%)  
        \\\cmidrule{2-4} \cmidrule{6-8}\cmidrule{10-12}\cmidrule{14-16}
        \rowcolor{LightCoreGreen}
        & Total & 968 & 137 (14\%) &  &
        137 (14\%) & 0 (0\%) & 137 (14\%) & & 
        158 (16\%) & 21 (2\%) & 342 (35\%) & & 
        177 (18\%) & 40 (4\%) & 647 (67\%)  
        \\ \midrule
        \textbf{ZC FL} & ZC FR & 242 & 54 (22\%) &  &
        54 (22\%) & 0 (0\%) & 54 (22\%) & &
        60 (25\%) & 6 (2\%) & 139 (57\%) & &
        65 (27\%) & 11 (5\%) & 232 (96\%)   
        \\
        & ZC RL & 242 & 59 (24\%) &  &
        59 (24\%) & 0 (0\%) & 59 (24\%) & &
        64 (26\%) & 5 (2\%) & 122 (50\%) & &
        68 (28\%) & 9 (4\%) & 192 (79\%) 
        \\
        & ZC RR & 242 & 37 (15\%) &  &
        37 (15\%) & 0 (0\%) & 37 (15\%) & &
        42 (17\%) & 5 (2\%) & 98 (40\%) & &
        46 (19\%) & 9 (4\%) & 213 (88\%)   
        \\
        & HPC I & 242 & 5 (2\%) &  &
        5 (2\%) & 0 (0\%) & 5 (2\%) & &
        5 (2\%) & 0 (0\%) & 5 (2\%) & &
        10 (4\%) & 5 (2\%) & 10 (4\%) 
        \\\cmidrule{2-4} \cmidrule{6-8}\cmidrule{10-12}\cmidrule{14-16}
        & Total & 968 & 155 (16\%) &  &
        155 (16\%) & 0 (0\%) & 155 (16\%) & & 
        171 (18\%) & 16 (2\%) & 364 (38\%) & & 
        189 (20\%) & 34 (4\%) & 647 (67\%)   
        \\ \midrule
        \textbf{ZC RL} & ZC FR & 242 & 49 (20\%) &  &
        49 (20\%) & 0 (0\%) & 49 (20\%) & &
        58 (24\%) & 9 (4\%) & 113 (47\%) & &
        65 (27\%) & 16 (7\%) & 232 (96\%) 
        \\
        & ZC FL & 242 & 39 (16\%) &  &
        39 (16\%) & 0 (0\%) & 39 (16\%) & &
        46 (19\%) & 7 (3\%) & 115 (48\%) & &
        53 (22\%) & 14 (6\%) & 232 (96\%)
        \\
        & ZC RR & 242 & 31 (13\%) &  &
        31 (13\%) & 0 (0\%) & 31 (13\%) & &
        43 (18\%) & 12 (5\%) & 107 (44\%) & &
        46 (19\%) & 15 (6\%) & 213 (88\%) 
        \\
        & HPC I & 242 & 4 (2\%) &  &
        4 (2\%) & 0 (0\%) & 4 (2\%) & &
        4 (2\%) & 0 (0\%) & 4 (2\%) & &
        10 (4\%) & 6 (2\%) & 10 (4\%)
        \\\cmidrule{2-4} \cmidrule{6-8}\cmidrule{10-12}\cmidrule{14-16}
        & Total & 968 & 123 (13\%) &  &
        123 (13\%) & 0 (0\%) & 123 (13\%) & & 
        151 (16\%) & 28 (3\%) & 339 (35\%) & & 
        174 (18\%) & 51 (5\%) & 687 (71\%) 
        \\ \midrule
        \textbf{ZC RR} & ZC FR & 242 & 50 (21\%) &  &
        50 (21\%) & 0 (0\%) & 50 (21\%) & &
        55 (23\%) & 5 (2\%) & 106 (44\%) & &
        65 (27\%) & 15 (6\%) & 232 (96\%) 
        \\
        & ZC FL & 242 & 41 (17\%) &  &
        41 (17\%) & 0 (0\%) & 41 (17\%) & &
        45 (19\%) & 4 (2\%) & 117 (48\%) & &
        53 (22\%) & 12 (5\%) & 232 (96\%) 
        \\
        & ZC RL & 242 & 39 (16\%) &  &
        39 (16\%) & 0 (0\%) & 39 (16\%) & &
        54 (22\%) & 15 (6\%) & 103 (43\%) & &
        68 (28\%) & 29 (12\%) & 192 (79\%) 
        \\
        & HPC I & 242 & 4 (2\%) &  &
        4 (2\%) & 0 (0\%) & 4 (2\%) & &
        4 (2\%) & 0 (0\%) & 4 (2\%) & &
        10 (4\%) & 6 (2\%) & 10 (4\%) 
        \\\cmidrule{2-4} \cmidrule{6-8}\cmidrule{10-12}\cmidrule{14-16}
        & Total & 968 & 134 (14\%) &  &
        134 (14\%) & 0 (0\%) & 134 (14\%) & & 
        158 (16\%) & 24 (2\%) & 330 (34\%) & & 
        196 (20\%) & 62 (6\%) & 666 (69\%) 
        \\ \midrule
        \textbf{HPC I} & ZC FR & 242 & 8 (3\%) &  &
        8 (3\%) & 0 (0\%) & 8 (3\%) & &
        19 (8\%) & 11 (5\%) & 26 (11\%) & &
        40 (17\%) & 32 (13\%) & 90 (37\%)
        \\
        & ZC FL & 242 & 11 (5\%) &  &
        11 (5\%) & 0 (0\%) & 11 (5\%) & &
        19 (8\%) & 8 (3\%) & 26 (11\%) & &
        53 (22\%) & 42 (17\%) & 232 (96\%) 
        \\
        & ZC RL & 242 & 8 (3\%) &  &
        8 (3\%) & 0 (0\%) & 8 (3\%) & &
        18 (7\%) & 10 (4\%) & 39 (16\%) & &
        51 (21\%) & 43 (18\%) & 131 (54\%)
        \\
        & ZC RR & 242 & 6 (2\%) &  &
        6 (2\%) & 0 (0\%) & 6 (2\%) & &
        14 (6\%) & 8 (3\%) & 16 (7\%) & &
        35 (14\%) & 29 (12\%) & 132 (55\%) 
        \\\cmidrule{2-4} \cmidrule{6-8}\cmidrule{10-12}\cmidrule{14-16}
        & Total & 968 & 33 (3\%) &  &
        33 (3\%) & 0 (0\%) & 33 (3\%) & & 
        70 (7\%) & 37 (4\%) & 107 (11\%) & & 
        179 (18\%) & 146 (15\%) & 585 (60\%)  
        \\ \midrule
        \rowcolor{LightCoreRed}
        \textbf{All} & Total & 4840 & 582 (12\%) &  &
        582 (12\%) & 0 (0\%) & 582 (12\%) & & 
        708 (15\%) & 126 (3\%) & 1482 (31\%) & & 
        915 (19\%) & 333 (7\%) & 3232 (67\%)
        \\ \bottomrule
    \end{tabular}
    \vspace{-10pt}
  \end{table*}

%% file: tex/5_conclusion.tex

\section{Conclusion and Outlook}%
\label{sec:conclusion_&_outlook}
In this work, we comparatively analyzed three strategies for integrating  separated control flows in a software-defined Ethernet backbone. We evaluated each separation concept in a real-world \ac{IVN}, which we transformed into a modern software-defined Ethernet topology.
The analysis showed that  embedding of exposed control flow information into the header fields can be matched by forwarding devices and improves network security and performance.
Hidden embeddings in application layer tunneling protocols have a significant oversupply of surplus control flows which  opens a security gap.

In detail, we could clearly show that exposing control flow identifiers to layer 2 packet headers  allows for precise identification of each vehicle control message and protect the information with a precise access control. Furthermore, a fine-grained control flow separation improves the precision in real-time configuration.
In contrast, application layer encapsulation using SOME/IP obfuscates control flows for \acp{IVN} and prevents a separate management of flows at the price of network security and performance.
A detailed case study confirmed that a precise and exposed control flow embedding allows for isolated communication contexts across the backbone and can establish trust zones between the gateways.

In future work, we will extend this analysis to cover a broader variety of in-vehicle communication flows and to advance the security in cars by additional network layer intelligence.

%% file: main.bbl
\begin{thebibliography}{10}
\providecommand{\url}[1]{#1}
\csname url@samestyle\endcsname
\providecommand{\newblock}{\relax}
\providecommand{\bibinfo}[2]{#2}
\providecommand{\BIBentrySTDinterwordspacing}{\spaceskip=0pt\relax}
\providecommand{\BIBentryALTinterwordstretchfactor}{4}
\providecommand{\BIBentryALTinterwordspacing}{\spaceskip=\fontdimen2\font plus
\BIBentryALTinterwordstretchfactor\fontdimen3\font minus
  \fontdimen4\font\relax}
\providecommand{\BIBforeignlanguage}[2]{{%
\expandafter\ifx\csname l@#1\endcsname\relax
\typeout{** WARNING: IEEEtran.bst: No hyphenation pattern has been}%
\typeout{** loaded for the language `#1'. Using the pattern for}%
\typeout{** the default language instead.}%
\else
\language=\csname l@#1\endcsname
\fi
#2}}
\providecommand{\BIBdecl}{\relax}
\BIBdecl

\bibitem{brkw-aeaJR-17}
S.~Brunner, J.~Roder, M.~Kucera, and T.~Waas, ``{Automotive E/E-Architecture
  Enhancements by Usage of Ethernet TSN},'' in \emph{2017 13th Workshop on
  Intelligent Solutions in Embedded Systems (WISES)}.\hskip 1em plus 0.5em
  minus 0.4em\relax Piscataway, NJ, USA: IEEE Press, Jun. 2017, pp. 9--13.

\bibitem{mabpp-oeiJR-08}
N.~McKeown, T.~Anderson, H.~Balakrishnan, G.~Parulkar, L.~Peterson, J.~Rexford,
  S.~Shenker, and J.~Turner, ``{OpenFlow: Enabling Innovation in Campus
  Networks},'' \emph{ACM SIGCOMM Computer Communication Review}, vol.~38,
  no.~2, pp. 69--74, 2008.

\bibitem{hmg-rsarn-18}
K.~Halba, C.~Mahmoudi, and E.~Griffor, ``{Robust Safety for Autonomous Vehicles
  through Reconfigurable Networking},'' in \emph{{Proceedings of the 2nd
  International Workshop on Safe Control of Autonomous Vehicles}}, ser.
  Electronic Proceedings in Theoretical Computer Science, vol. 269.\hskip 1em
  plus 0.5em minus 0.4em\relax Open Publishing Association, 2018, pp. 48--58.

\bibitem{krst-gseJR-11}
A.~Kern, D.~Reinhard, T.~Streichert, and J.~Teich, ``{Gateway Strategies for
  Embedding of Automotive {CAN}-Frames into Ethernet-Packets and Vice Versa},''
  in \emph{Architecture of Computing Systems - {ARCS} 2011}.\hskip 1em plus
  0.5em minus 0.4em\relax Springer Berlin Heidelberg, 2011, pp. 259--270.

\bibitem{a-somip-16}
AUTOSAR, ``{SOME/IP Protocol Specification},'' AUTOSAR, AUTOSAR Standard 696,
  Nov. 2016.

\bibitem{hrwh-rtcJR-15}
C.~Herber, A.~Richter, T.~Wild, and A.~Herkersdorf, ``{Real-Time Capable {CAN}
  to {AVB} Ethernet Gateway using Frame Aggregation and Scheduling},'' in
  \emph{Design, Automation {\&} Test in Europe Conference {\&} Exhibition
  ({DATE}), 2015}.\hskip 1em plus 0.5em minus 0.4em\relax Piscataway, NJ, USA:
  {IEEE} Press, 2015.

\bibitem{rdgks-iabJR-19}
M.~Rumez, A.~Duda, P.~Grunder, R.~Kriesten, and E.~Sax, ``{Integration of
  Attribute-based Access Control into Automotive Architectures},'' in
  \emph{2019 {IEEE} Intelligent Vehicles Symposium ({IV})}.\hskip 1em plus
  0.5em minus 0.4em\relax Piscataway, NJ, USA: {IEEE} Press, jun 2019.

\bibitem{wlk-ncaJR-19}
J.~Wang, J.~Liu, and N.~Kato, ``{Networking and Communications in Autonomous
  Driving: A Survey},'' \emph{{IEEE} Communications Surveys {\&} Tutorials},
  vol.~21, no.~2, pp. 1243--1274, 2019.

\bibitem{hl-rscJR-18}
Q.~Hu and F.~Luo, ``{Review of Secure Communication Approaches for In-Vehicle
  Network},'' \emph{International Journal of Automotive Technology}, vol.~19,
  no.~5, pp. 879--894, sep 2018.

\bibitem{krvra-sdncs-15}
D.~Kreutz, F.~M.~V. Ramos, P.~E. Ver\'{i}ssimo, C.~E. Rothenberg,
  S.~Azodolmolky, and S.~Uhlig, ``{Software-Defined Networking: A Comprehensive
  Survey},'' \emph{Proceedings of the IEEE}, vol. 103, no.~1, pp. 14--76, Jan.
  2015.

\bibitem{sxhg-ensJR-16}
S.~Shin, L.~Xu, S.~Hong, and G.~Gu, ``{Enhancing Network Security through
  Software Defined Networking ({SDN})},'' in \emph{2016 25th International
  Conference on Computer Communication and Networks ({ICCCN})}.\hskip 1em plus
  0.5em minus 0.4em\relax Piscataway, NJ, USA: {IEEE} Press, aug 2016.

\bibitem{hmks-snsti-19}
T.~H{\"a}ckel, P.~Meyer, F.~Korf, and T.~C. Schmidt, ``{Software-Defined
  Networks Supporting Time-Sensitive In-Vehicular Communication},'' in
  \emph{2019 IEEE 89th Vehicular Technology Conference (VTC2019-Spring)}.\hskip
  1em plus 0.5em minus 0.4em\relax Piscataway, NJ, USA: IEEE Press, Apr. 2019,
  pp. 1--5.

\bibitem{wjscn-kzdJR-14}
L.~Wang, S.~Jajodia, A.~Singhal, P.~Cheng, and S.~Noel, ``k-Zero Day Safety: A Network Security Metric for Measuring the Risk of Unknown Vulnerabilities,''
  \emph{{IEEE} Transactions on Dependable and Secure Computing}, vol.~11,
  no.~1, pp. 30--44, jan 2014.

\bibitem{rwwir-sraJR-09}
A.~Ruddle, D.~Ward, B.~Weyl, S.~Idrees, Y.~Roudier, M.~Friedewald, T.~Leimbach,
  A.~Fuchs, S.~G\"urgens, O.~Henniger, R.~Rieke, M.~Ritscher, H.~Broberg,
  L.~Apvrille, R.~Pacalet, and G.~Pedroza, ``\BIBforeignlanguage{en}{{Security
  Requirements For Automotive On-Board Networks Based On Dark-Side
  Scenarios}},'' \emph{\BIBforeignlanguage{en}{Evita Deliverable 2.3}}, 2009.

\bibitem{cmkas-ceaas-11}
S.~Checkoway, D.~Mccoy, B.~Kantor, D.~Anderson, H.~Shacham, S.~Savage,
  K.~Koscher, A.~Czeskis, F.~Roesner, and T.~Kohno, ``{Comprehensive
  Experimental Analyses of Automotive Attack Surfaces},'' in \emph{Proceedings
  of the 20th USENIX Security Symposium}, vol.~4.\hskip 1em plus 0.5em minus
  0.4em\relax USENIX Association, Aug. 2011, pp. 77--92.

\bibitem{mv-sraJR-14}
C.~Miller and C.~Valasek, ``{A Survey of Remote Automotive Attack Surfaces},''
  \emph{black hat USA}, vol. 2014, 2014.

\bibitem{mv-reuJR-15}
------, ``{Remote Exploitation of an Unaltered Passenger Vehicle},''
  \emph{Black Hat USA}, vol. 2015, p.~91, 2015.

\bibitem{lccz-sdfJR-19}
S.~Longari, A.~Cannizzo, M.~Carminati, and S.~Zanero, ``A Secure-by-Design Framework for Automotive On-board Network Risk Analysis,'' in \emph{2019
  {IEEE} Vehicular Networking Conference ({VNC})}.\hskip 1em plus 0.5em minus
  0.4em\relax Piscataway, NJ, USA: {IEEE} Press, dec 2019.

\end{thebibliography}
